\documentclass{emulateapj}

\shorttitle{Neutral Winds in Sunspots}
\shortauthors{Kuhn et al.}

\begin{document}

\title{Are Neutral Sunspot Winds Important for Penumbral Dynamics and the First
Ionization Potential Effect?}

\author{J. R. Kuhn, H. Lin, H. Morgan}

\affil{Institute for Astronomy, University of Hawaii, Honolulu-HI-96822}

\begin{abstract}

The low ionization state in  parts of a sunspot may play an
important role in its evolution and dynamical state. The cool magnetic
interior region of the sunspot develops a substantial neutral atomic and
molecular hydrogen osmotic pressure which can drive a wind outward from
the umbra.  Ambipolar diffusion against the magnetically pinned ionized
plasma component can also distort the umbral magnetic field into a
filamentary penumbral structure. This may be important for explaining the
development of the sunspot penumbra and the Evershed flow.
This fractionation process may also be important for
the ``First Ionization Potential'' (FIP) effect seen in the solar wind.
In support of this mechanism we find evidence for such ionization
fractionization in UV observations of molecular hydrogen in a sunspot
umbra and penumbra.

\end{abstract}


\section{Introduction}

Existing sunspot models fall short in explaining all their observational
facts. One puzzling dynamical feature has been the outward
Evershed flow (cf. Bray and Loughhead 1962). To account for this,
Thomas and Weiss (1992) developed models based on flux-tube siphon
flows. Jahn and Schmidt (1994) accounted for these flows by treating
the penumbra in terms of convective interchange mitigated by whole flux
tubes. Recently, high resolution observations of sunspot penumbrae,
like those of Scharmer et al.  (2002) and Rimmele (2004), have shown
that fine structure in sunspot penumbra is more complex than either of
these pictures directly accounts for.  Our discussion here 
illustrates another physical concept that may help to explain the
penumbral dynamics now being observed.

A region with a sharply bounded magnetic field in a partially ionized,
thermally differentiated, plasma may generate dynamically important {\it
cross-field} flows. We consider an idealized representation of the
umbral magnetic field near the temperature minimum in a sunspot using a
two-fluid model consisting of a dominant neutral and a tenuous ionized
plasma component. As is observed in sunspot umbrae (prior to the
appearance of penumbrae) there is a substantial temperature gradient
between the inner (magnetized) and outer (ionized) plasma. In the Sun
the cooler inner region is insulated from the hotter atmosphere (near
the photospheric level) by the magnetic field, which prevents
convective energy penetration from the external atmosphere, and by the
relatively opaque  $H^-$-rich photosphere which provides radiative
isolation from the nearby hot gas.  A consequence of this
stratification is a strong horizontal gradient in the density of
neutral and molecular hydrogen and the temperature between the magnetized and photospheric
plasma. The diffusive (osmotic) pressure associated with this gradient
and the resulting flow may have important consequences for the dynamics
of a penumbral region, the outward Evershed flow, and the FIP effect in
the solar wind.

\section{A two component plasma near the temperature minimum}

At unity optical depth
(i.e. reference height $z=0$ with positive height measured outward),
and in a plasma with a  temperature of about 4000K, as in Zwaan's (1974)
sunspot models, the ratio of the electronic partial pressure to the sum of
neutral H and molecular H$_2$ is about $10^{-5}.$  More recent models
(Maltby et al. 1986, Pizzo 1988, Collados et al. 1994) are not qualitatively
different and suggest that umbral temperatures near $z=0$
may be closer to 3000K, perhaps with an even smaller ionization
fraction.  These and equilibrium one-dimensional sunspot models (cf.
Fontenla et al. 1999) suggest that this low temperature ``neutral zone'' 
extends upward to about $z=500km.$  

Near $z=0$ realistic numerical models of the upper convection zone and
photosphere (Stein and Nordlund 1998) show that temperature falls
steeply with height due to vanishing $H^-$ opacity as the plasma
electron density begins to drop. This region is about 500km below the
mean atmosphere temperature minimum. Beneath this rapid temperature
drop the $H^-$ opacity dominates visible and IR wavelengths. The
sunspot temperature minimum region is a few hundred kilometers below
the mean atmosphere temperature minimum, near the top of the convection
zone. At this height the horizontal gradients in temperature and
electron density between the magnetic umbra and the non-magnetic
photosphere are large.  A useful classical solution for the field and
temperature geometry of an umbral flux tube is calculated by Pizzo
(1986), yielding a Wilson depression of typically 500km and a core to
exterior temperature gradient near the flux tube temperature minimum of
$\approx$5000K.

The cold umbra has a considerably lower electron density and its much
lower $H^-$  opacity allows this non-convective umbral plasma to
radiate upward and to cool. The mean atmosphere models of Fontenla et
al. 1999 also illustrate these differences. Estimates of the mean $H^-$
density imply, at visible and near-IR wavelengths and near $z=0$, that
the photon free-path is short enough that the interior hot wall of the
sunspot umbra is radiatively isolated from the cool sunspot core (cf.
Joshi, Punetha, and Pande, 1979).

In general we expect the umbral region of a sunspot, before the
development of its penumbra, to be several thousand degrees cooler than
the surrounding, relatively $H^-$-rich, photosphere. In a few-$100km$ region
below the temperature minimum the umbra is primarily neutral gas with
an ionization fraction of $10^{-5}$ or less and (from Fontenla et al.
1999 models) a hydrogen mass density that ranges from about $10^{-7}$
to $10^{-10}g/cm^3$.

\section{A neutral wind and leaky flux regions}

In hydrostatic and magnetic equilibrium we expect the gas and magnetic
pressure within the umbra to equal the exterior pressure.
In the neutral zone most of the umbral gas does not interact with the
field so that only $B$ contributes to the interior force balance. Under these conditions the neutral gas is free to ``leak"
from the high field umbral region into the surrounding hot photosphere.
Since the neutral fluid density is low outside the umbra this osmotic
pressure can be significant -- essentially equal to the initial gas
pressure in the magnetized region.  The only impediment to this outward
flow is the net collisional momentum transfer to the neutral atoms and
molecules from the tenuous ionized component of the plasma which
remains tightly coupled to the magnetic field.

The dynamics of a neutral and ionized fluid in a magnetic field (closely
related to ambipolar diffusion), has been considered in the solar atmosphere before
(cf. Fontenla et al. 1990, Schmieder et al. 1999, Chitre and Krishan
2001) but, to our knowledge, has not been applied to sunspots.  Here we formulate the problem in terms of an ionized plasma
of density $\rho_i=\rho f$, a neutral plasma $\rho_n=\rho (1-f)$, the
respective neutral and ionized plasma bulk velocity fields (${\bf
v_{i,n}}$), pressures ($p_{i,n}$), magnetic field (${\bf B}$), and
collision rate between neutrals and ions ($\gamma_{ni}$) and ions
against neutrals ($\gamma_{in}$). In this case the force density on
each fluid component can be expressed as $$\rho_i{d{\bf v_i}\over
dt}={({\bf\nabla}\times{\bf B})\times{\bf B}\over 4\pi}
-{\bf\nabla}p_i-\gamma_{in}({\bf v_i-v_n})\rho_i\eqno{(1)},$$ and
$$\rho_n{d{\bf v_n}\over dt}= -{\bf\nabla}p_n-\gamma_{ni}({\bf
v_n-v_i})\rho_n\eqno{(2)}.$$ In our idealized problem we consider a
vertical B field ($z$-direction) that is non-zero to the left (at
$x<0$) and zero at $x>0$. The relative bulk separation velocity between
ions and neutrals (``slip'' velocity) is given by ${\bf v_i}-{\bf
v_n}$.  From  equation (1) with no slip velocity we obtain simply
$-\delta B^2/8\pi = \delta p_i$ the magnetostatic pressure balance
condition across the magnetic interface. We assume the ions are pinned
to a fixed magnetic field. In equation (2), if we have ${\bf v_i}=0$ then
in a steady-state we obtain $dp_n/dx=-\gamma_{ni}\rho_n{\bf
v_n}$.

The momentum ``drag'' transfer to the neutral fluid is dominated by the
ion collision rate which at low velocities,
corresponding to Evershed flows of less that a few $km/s$, is
independent of the slip velocity  (Draine et al. 1983). In this case
$\gamma_{ni}\approx 1\times 10^{-9}n_e~[cm^3/s]$ where $n_e$ is the ion
number density and we've assumed ions and neutrals have similar masses.
Initially the interior neutral gas pressure is, plausibly, comparable to the
magnetic pressure ($p_n\approx B^2/8\pi$) and we take $l$ to describe
the magnetic boundary thickness. The neutral ``wind'' velocity across
the magnetic boundary is then $$v_n = 7\times 10^{-17} {B^2\over
\rho^2fl} [cm/s]\eqno{(3)}$$ where $f$ is the ionization fraction. Here magnetic
field, density and boundary thickness are measured in Gauss and cgs
units.

The neutral wind velocity depends sensitively on the thermal and magnetic 
boundary thickness.  
The magnetic boundary thickness may be quite small since the exterior
convection zone  concentrates  stray flux into the downdraft
regions immediately surrounding the umbra (wherein there is no convective
flow). Observations from Scharmer et al. (2002)
achieve a spatial resolution of 90km but do not resolve the magnetic penumbral fine structure. The thermal stratification is dominated by
H$^-$ opacity and, based on the molecular models of sunspots by
Joshi et al. (1979), near unity optical depth (at 500nm) the radiation
length (1/$\kappa\rho$) can be less than 10km. Thus the flux concentrating
effect of surrounding convection and the strong temperature dependence of
$H^-$ opacity suggests to us that the magnetic/thermal boundary may be
quite sharp. Lacking observational constraints we assume here a value of 10km, but these estimates may change if we learn more from the next generation of high resolution
observations. With $l=10km$ in the deep neutral zone,  $\rho =
10^{-7}g/cm$, $B=3kG$, and $f=10^{-5}$ then $v_n \approx 63 m/s$. We
note that the neutral wind velocity scales approximately as $ p/\rho^2$
so it is a steep function of vertical position in the atmosphere,
increasing approximately exponentially with height in the neutral zone
until the temperature and ionization fraction are large enough to
quench the osmotic flow.

Unlike ordinary diffusion flows, the neutral wind escaping from the
umbra  is quickly ionized when it penetrates the magnetic interface.
Consequently there is no ``back-filling'' neutral concentration at
$x>0$ to suppress the flow, although it halts immediately beyond the
magnetic boundary in the radiative/convective exterior where it mixes with the turbulent plasma.

\section{Penumbra Dynamics}

Penumbrae develop from large pores and umbral regions.  These filamentary (but sharply
defined) structures eminate from the umbra over a typical timescale of
about a day and with a typical radial extent of about 7500km (cf.
Bray and Loughhead, 1965). According to Zwaan (1992) the penumbra seems to
develop at the expense of the umbral magnetic flux. If there is a
characteristic velocity with which the penumbra evolves it is not
fast, perhaps 50-100 $m/s$. Interestingly, this is the velocity scale of the
neutral wind in the deeper neutral zone. We note that
the observed scale of the outward velocity
of the Evershed flow in the penumbra is typically 1$km/s$ or larger (cf. Thomas and Weiss, 1992).

The neutral wind creates a non-zero third term on the
RHS of Eq. 1 which implies a positive acceleration of the ionic fluid.
Thus this wind drags the ions and the magnetic field outward away from
the umbra.  

The steady-state solution to Eqns. (1) and (2) is $\delta p_n+\delta
p_i = -B^2/8\pi$. The magnetized region is carried
outward until the {\it total} neutral and ionic pressure difference
across the magnetic interface balances the magnetic pressure. Note,
that as long as the magnetized plasma remains thermally
isolated from the exterior (with a large neutral component) and the
external ionized pressure doesn't increase then there can be
{\it no time-independent solution}. In a real
penumbra equilibrium should occur when: 1) the neutral pressure is sufficiently
small, either because the penumbral filament is no longer effectively
thermally isolated from the hot photosphere, or 2) the neutral wind
successfully evacuates the interior field region, or 3) the magnetic field
is small enough.

\section{Some features and expectations of a neutral wind sunspot model}

While our description is far from a complete model it has
some attractive features which warrant further study:

In this picture a penumbral filament (``flux region'') 
is advected from the umbra into
the photosphere by a vertically localized horizontal outward flow. The
penumbra forms near the top of the convectively unstable exterior.
Figure 1 shows a cartoon representation of this.

The neutral wind is driven by the umbra/photosphere and moving
penumbra/photosphere interface. We therefore expect the flow along a
penumbral filament to include the effects of: 1) the continuous cross-field
leakage of neutral gas all along the length of the filament/photosphere
interface, and 2) a parallel ionized and neutral outward flow channeled
along the field lines of the
horizontally elongated penumbral flux region.

\begin{figure}
\includegraphics[width=\linewidth]{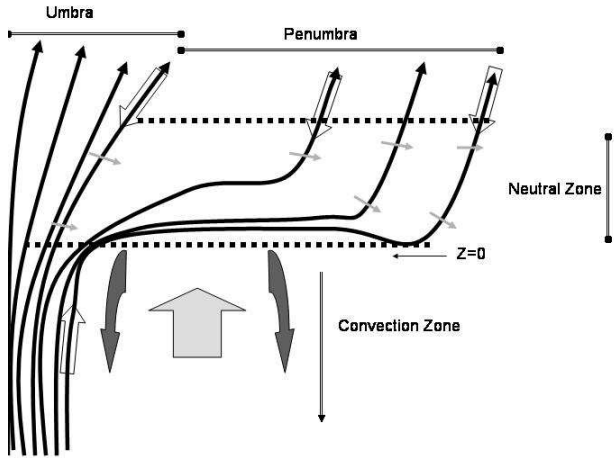}
\figcaption[cmd.eps]{\label{fig:cmd}
An umbral region (left) with a primarily vertical B field (dark lines)
has its magnetic field advected into the ``neutral zone'' by the
cross-field neutral wind (short grey arrows). Entrained ionized and
neutral plasma flows along field lines from above and below to
replenish the low temperature sunspot plasma. Transparent arrows
indicate deep upflows and possible reverse Evershed flows corresponding
to the upper level downflows.}
\vspace*{-7mm}
\end{figure}

The neutral plasma which evacuates the umbral and penumbral flux regions may
be replaced by mass upflows from below and by downflows from the flux 
regions that connect with the hotter upper regions of the chromosphere
and transition region/corona. These downflows might be 
the siphon-like reverse Evershed flows observed in the chromosphere and
above (cf. Georgakilas and Christopoulou, 2003).

Although the magnetic interface-driven neutral flow is 10-100m/s the observed
outward Evershed flow can be one or two orders of magnitude larger.
How can such a large velocity be a consequence of the osmotic
flow? The partially ionized plasma flow in the penumbra will be
channeled by the nearly horizontal magnetic field. Unlike the umbra,
the penumbra is filamentary so that the neutral outflow
exists along the length of the filament. Thus, within the filament the parallel flow
velocity along the field lines must be larger than the wind
velocity by a factor of approximately the ratio of the effective area
of the filament to its cross-section. In detail this is difficult to
estimate but from the Scharmer et al. (2002) high resolution images the
length-to-diameter ratio of the filaments can easily be 10, yielding
channeled flow velocities that are two orders of magnitude larger than
the interface neutral wind velocity.

\section{Ionic fractionization, FIP effect}

An unusual feature of this model is that it tends to predict a larger
charged-species concentration in the umbra than in the penumbra. Conversely we
expect a larger concentration of neutrals in the penumbra as they
are siphoned from the umbra. For many molecular and atomic
species this tendency can be masked by local ionization
equilibrium effects caused by the temperature difference between the
umbra and penumbra, but in general we expect elements with a $lower$ first
ionization potential to have a $higher$ density in the strong field
region. This can have interesting consequences for the
solar wind FIP effect if the wind acceleration mechanism is
related to strong B-field regions deep in the chromosphere
or even the photosphere. We note that wind observations show that
low first ionization elements are most strongly enhanced (a factor of 3-5)
in low latitude, active-region, slow solar winds (McKenzie and Feldman, 1992). These are precisely
where we expect underlying sunspots to cause enhanced ionized
species abundances in the strong field regions.

\section{Evidence of ionic fractionization in sunspots}

It may be possible to detect a superabundance of neutral atoms or
molecules in the penumbra, although this may be masked by the effects
of the relatively short vertical optical path through the penumbra (as
compared to the umbra -- see Fig. 1) and by the details of the line
formation and strong temperature dependence of the molecular
dissociation equilibrium abundances. While there are few observations of
molecular hydrogen, interpreting its spectra may be more straightforward
because it is seen in emission in the far ultraviolet part of the
spectrum.  These lines are likely to be optically thin and it has a
monotonically increasing molecular density with decreasing temperature
in the photosphere (cf. Aller 1963).

\begin{figure}
\includegraphics[width=\linewidth]{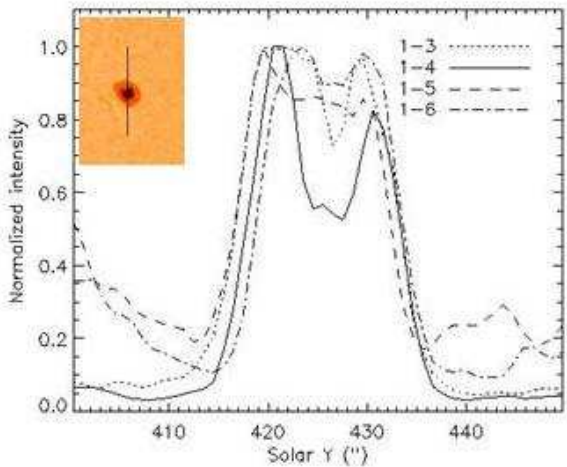}
\figcaption[sumerh2.ps]{\label{fig:sum}
Inset - MDI/SOHO intensity image showing the leading sunspot of active region NOAA 8487 observed on 1999 March 18. The vertical black line shows the position of the SUMER slit across the sunspot. Intensities of several ultraviolet H$_2$ spectral lines across the sunspot as measured by the SUMER/SOHO instrument on 1999 March 18. The legend refers to the Werner band of each line. For clarity, the intensities are spatially smoothed with a narrow Gaussian of FWHM 3 spatial bins ($\sim$3") and are normalized to unity.
}
\vspace*{-7mm}
\end{figure}

According to Zwaan (1974), H$_2$ can account for 10\% or more of the
atomic H partial pressure in the sunspot neutral zone.  Molecular
hydrogen has 8 prominent Werner band transmissions (1-0 to 1-7) which
emit lines within the wavelength range of SUMER.  These are listed by
Bartoe et al. (1979).  A previous analysis of the H$_2$ lines in this
sunspot were made by Schuhle et al. (1999).  In the SUMER data, the
1-0, 1-1 and 1-7 Werner band lines are heavily blended with emission
lines of various ions and the 1-2 line has a very low intensity.  These
are disregarded for this study.  The remaining four Werner band lines
(1-3 to 1-6) have theoretical central wavelengths of 1119.08, 1163.81,
1208.94 and 1254.12\AA\ respectively.  The lines are easily identified
in the SUMER sunspot observations. The line intensity has been
integrated over the wavelength range of each line in each spatial bin.
Figure 3 shows the SUMER observations in context with an inset image of
the sunspot as observed by the Michelson Doppler Imager (MDI).  This
figure shows the normalized intensity of the four Werner bands across
the region of the SUMER FOV containing the sunspot.  Differences
between the spatial profiles of the bands may be due to blends or
temporal changes within the sunspot, although these four observations
were all made within one hour.  All four bands show their highest
intensity within the sunspot penumbra and a sharp decrease in intensity
at the photosphere.  The striking decrement in H$_2$ intensity in the
umbra is obvious.  This is most prominent in the 1-4 band line.

Jordan et al. (1978) discovered and analyzed many H$_2$ UV emission
lines to conclude that they were due to {\it flourescence} from hotter
transition region UV line radiation illuminating the molecular hydrogen
from above. A fully self-consistent line-formation model of these
Werner band measurements is beyond the scope of this letter but it is
important to note that Jordan et al. (1978) concluded that these UV
lines should be optically thin. Because the equilibrium number density
of molecular hydrogen monotonically rises with decreasing
temperature (cf. Aller 1963) it is quite possible that the weaker H$_2$ line
intensity from the cooler umbra is due to a significantly lower umbral
than penumbral H$_2$ number density.

\section{Conclusions}

The radiatively isolated, cool,  sunspot umbra should develop a
significant osmotic pressure which drives a neutral wind outward into
the surrounding photosphere and penumbra. Typical velocities can be
10-100$m/s$ and given expected umbral ionization fractions these
velocities will carry some of the umbral magnetic flux with them. This
neutral wind tends to evacuate the high-field umbral region of its
neutral atomic and molecular gas. A likely consequence of this is
upflows from below and a reverse Evershed flow from the chromosphere
above in order to maintain mass conservation. It seems difficult to
avoid such a neutral wind in a  partially ionized region where there
is a common magnetic and temperature boundary.  Observations of
molecular H$_2$ in sunspots tend to confirm that neutral gas is
superabundant in the penumbra in comparison to the umbra. If the slow
solar wind ultimately originates from high field regions near the
photosphere this neutral wind might also lead to a low first ionization
potential (FIP) element enhancement in the solar wind.

\acknowledgements
This research was supported under grants from the NASA SRT and the
Air Force Multidisciplinary University Research Initiative (MURI)
programs. We're grateful to an anonymous referee for improving this
discussion.


\begin{thebibliography}{}

\bibitem[all]{all}Aller, L. H. 1963, {\it The Atmospheres of the Sun and Stars}
(2nd Ed.; New York: Ronald Press)
\bibitem[bar]{bar}Bartoe, J.D., Brueckner, G., Nicolas, K., Sandlin, G., VanHoosier, M., Jordon, C., 1979, \mnras, 187, 463
\bibitem[bra]{bra}Bray, R.J., Loughhead, R. {\it Sunspots} (Wiley Inc, NY) 1962
\bibitem[chi]{chi}Chitre, S.M., and Krishan, V., 2001, \mnras, 323, 23
\bibitem[col]{col}Collados, M., Martinez-Pillet, V., Ruiz Cobo, R., del Toro Iniesta, J. C., Vazquez, M. 1994, Astron. Astrophys., 291, 622
\bibitem[Curdt et al.(2001)]{cur2001} Curdt, W., Brekke, P., 
Feldman, U., Wilhelm, K., Dwivedi, B.~N., Sch{\" u}hle, U., \& Lemaire, P.\ 
2001, \aap, 375, 591 
\bibitem[dra]{dra}Draine, B.T., Roberge, W.G., Dalgarno, A., 1983, \apj, 264, 485
\bibitem[fon2]{fon2}Fontenla, J. M., Avrett, E.H., Loeser, R., 1990 \apj, 355, 700
\bibitem[fon]{fon}Fontenla, J. M., White, O.R., Fox, P.A., Avrett, E.H., and Kurucz, R. L. 1999, \apj, 518, 480
\bibitem[geo]{geo}Georgakilas, A.A., Christopoulou, E.B. 2003, \apj, 584, 509
\bibitem[jah]{jah}Jahn, K., Schmidt, H.U. 1994, Astr.Astrophys., 290, 295
\bibitem[jor]{jor}Jordan, C., Brueckner, G., Bartoe, J-D., Sandlin, G.D., VanHoosier, M., 1978, \apj, 226, 687
\bibitem[jos]{jos}Joshi, G.C., Punetha, L.M., and Pande, M.C., 1979, Sol. Phys., 64, 255
\bibitem[mal]{mal}Maltby, P., Avrett, E.H., Carlsson, M., Kjeldseth-Moe, O., Kurucz, R.L., Loeser, R., 1986, \apj, 306, 284
\bibitem[mck]{mck}McKenzie, D.L., Feldman, U. 1992, \apj, 389, 764
\bibitem[piz]{piz}Pizzo, V. 1986, \apj, 302, 785
\bibitem[rim]{rim}Rimmele, T. 2004, \apj, 604, 906
\bibitem[sch]{sch}Scharmer, G.B., Gudiksen, B.V., Kiselman, D., Lofdahl, M.G., Rouppe van der Voort, L. 2002, Nature, 420 151
\bibitem[smi]{smi}Schmieder, B., Heinzel, P., Vial, J.C., Rudawy, P., 1999, Solar Phys., 189, 109
\bibitem[schu]{schu}Schuehle, U., Brown, C.M, Curdt, W., Feldman, U., 1999,
in 8the SOHO Workshop: Plasma Dynamics and Diagnostics in the Solar Transition
Region and Corona, ed. J. C. Vial, and B. Kaldeich-Schumann
(ESA publ. no. 446: Paris), 617
\bibitem[nor]{nor}Stein, R.F., and Nordlund, A., 1998, \apj, 499, 914
\bibitem[tho]{tho}Thomas, J.N., Weiss, N. 1992, in Sunspots: Theory and
Observations, ed. Thomas, J. and Weiss, N. (NATO ASI Ser. C, 375; Dordrecht: Kluwer), 3
\bibitem[Wilhelm et al.(1995)]{wil1995} Wilhelm, K., et al.\ 
1995, \solphys, 162, 189
\bibitem[zwa]{zwa}Zwaan, C. 1974, Solar Phys., 37, 99
\bibitem[zwa2]{zwa2}Zwaan, C. 1992, in Sunspots: Theory and
Observations, ed. Thomas, J. and Weiss, N. (NATO ASI Ser. C, 375; Dordrecht: Kluwer), 75

\end{thebibliography}
\end{document}